\documentclass[fleqn,usenatbib]{mnras}

\usepackage[T1]{fontenc}
\usepackage{newtxtext,newtxmath}

\DeclareRobustCommand{\VAN}[3]{#2}
\let\VANthebibliography\thebibliography
\def\thebibliography{\DeclareRobustCommand{\VAN}[3]{##3}\VANthebibliography}

\usepackage{graphicx}	
\usepackage{amsmath}	
\usepackage[flushleft]{threeparttable}

\title[1A 0535+262 during the giant outburst in 2020]{Study of timing and spectral properties of the X-ray pulsar 1A 0535+262 during the giant outburst in 2020 November-December}

\author[M. Mandal \& S. Pal]{
Manoj Mandal$^{1}$,
Sabyasachi Pal$^{1}$\thanks{E-mail: sabya.pal@gmail.com}
\\
$^{1}$Midnapore City College, Kuturia, Bhadutala, West Bengal, India 721129\\
}

\date{Accepted 2022 January 11. Received 2022 January 11; in original form 2021 April 23}

\pubyear{2021}

\begin{document}
\label{firstpage}
\pagerange{\pageref{firstpage}--\pageref{lastpage}}
\maketitle


\begin{abstract}
We made a detailed study of the timing and spectral properties of the X-ray pulsar 1A 0535+262 during the recent giant outburst in 2020 November and December. The flux of the pulsar reached a record value of $\sim$12.5 Crab as observed by {\it Swift}/BAT (15--50 keV) and the corresponding mass accretion rate was $\sim6.67\times10^{17}$ g s$^{-1}$ near the peak of the outburst. There was a transition from the subcritical to the supercritical accretion regime which allows exploring different properties of the source in the supercritical regime. A q-like feature was detected in the hardness-intensity diagram during the outburst. We observed high variability and strong energy dependence of pulse profiles during the outburst. Cyclotron Resonant Scattering Feature (CRSF) was detected at $\sim44$ keV from the {\it NuSTAR} energy spectrum in the subcritical regime and the corresponding magnetic field was $B\simeq4.9\times10^{12}$ G. The energy of the CRSF was shifted towards lower energy in the supercritical regime. The luminosity dependence of the CRSF was studied and during the supercritical regime, a negative correlation was observed between the line energy and luminosity. The critical luminosity was $\sim6\times10^{37}$ erg s$^{-1}$ above which a state transition occurred. A reversal of correlation between the photon index and luminosity was observed near the critical luminosity. The {\it NuSTAR} spectra can be described by a composite model with two continuum components, a blackbody emission, cut-off power law, and a discrete component to account for the iron emission line at 6.4 keV. An additional cyclotron absorption feature was included in the model.

\end{abstract}


\section{Introduction}
\label{intro}
The X-ray pulsar 1A 0535+262 was discovered by the {\it Ariel} V spacecraft in 1975 April during a giant outburst with a period of 104 s \citep{Co75, Ro75} and an orbital period of $P_{orb}=111.1\pm0.3$ d \citep{Fi96}. It is a pulsating X-ray binary system associated with the Be star HDE 245770 \citep{Li79} in a highly eccentric orbit with eccentricity $e=0.42\pm0.02$ \citep{Fi96}. The source went through several giant outbursts in 1975, 1977, 1980, 1983, 1989, 1994, 2005, 2009, and 2011 \citep{Na82,Ma89,Se90,Wi94,Fi96,Ca11}.  
During the 1994 outburst, Quasi-Periodic Oscillations (QPOs) were observed within the range 27--72 mHz \citep{Fi96}, with frequency correlated to the X-ray flux and spin-up torque \citep{Fi96, Gh96}. The source went through a giant outburst again in 2020 November after $\sim$10 years.

 Critical luminosity ($L_{\textrm{crit}}$) is the luminosity above which a state transition from subcritical to supercritical takes place. The subcritical state ($L_{X}<L_{\textrm{crit}}$) is known to be the low luminosity state whereas the supercritical state is high luminosity state ($L_{X}>L_{\textrm{crit}}$) \citep{Be12}. The critical luminosity is crucial to determine whether the radiation pressure of the emitting plasma is capable of decelerating the accretion flow \citep{Ba76, Be12}. The luminosity during the 2020 giant outburst reached a record high, which was significantly higher than the critical luminosity \citep{Re13}. The source entered to a supercritical regime from a subcritical regime during the outburst. In the supercritical regime, radiation pressure is high enough to stop the accreting matter at a distance above the neutron star, forming a radiation-dominated shock \citep{Ba76, Be12}. For the subcritical regime, accreted material reaches the neutron star surface through nuclear collisions with atmospheric protons or through Coulomb collision with thermal electrons \citep{Ha94}. These accretion regimes can also be probed by noting changes in the cyclotron line energies, pulse profiles, and changes in the spectral shape \citep{Pa89, Re13}.
 During the transition from the subcritical to the supercritical regime, sources show two different branches in their hardness-intensity diagram (HID) which are known as horizontal branch (HB) and diagonal branch (DB) \citep{Re13}. The HB implies the low-luminosity state of the source, which is represented by spectral changes and high X-ray variability. The DB corresponds to the high-luminosity state that appears when the X-ray luminosity is above the critical limit. The classification HB and DB depends on HID patterns that the source follows. The HB pattern is generally observed in the subcritical regime and the DB pattern is observed in the supercritical regime \citep{Re13}.
 
 We focused on the recent giant outburst of 1A 0535+262 detected by Swift/BAT\footnote{\url{https://swift.gsfc.nasa.gov/results/transients/}}, {\it MAXI}/GSC\footnote{\url{http://maxi.riken.jp/top/index.html}}, and {\it Fermi}/GBM in 2020 November \citep{Ja20, Ma20, Na20, Pa20}.   
 
The outburst continued for a long time (nearly seven weeks) and was observed by different satellites ({\it NuSTAR}, {\it Swift}, {\it NICER}, {\it Fermi}, and {\it Chandra}). The source was also monitored using {\it Insight}-HXMT during the outburst \citep{Co21}. We have studied the evolution of different timing and spectral properties of the source during the outburst using {\it NuSTAR} and {\it NICER} observations. 
 
  Earlier, a cyclotron line near 50 keV and a harmonic around 100 keV were detected for this X-ray pulsar \citep{Ke94, Gr95, Kr96, Ts19}.
The magnetic field of a neutron star can be simply determined using Cyclotron Resonant Scattering Feature (CRSF). The magnetic field strength is directly proportional to the energy of the fundamental line and the spacing between harmonics \citep{Mu13}. 

 In this work, we investigated the dependence of the spectral and timing parameters with luminosity and focused on the evolution of the pulse profile, pulsed fraction, beaming pattern, and emission geometry as the source transited from the subcritical to the supercritical accretion regime.
The data reduction and analysis methods are discussed in Section \ref{obs}. We have summarized the result of the current study in Section \ref{res}. Discussion and conclusion are summarized in the Section \ref{dis} and \ref{con}, respectively.

 \begin{figure}
\centering{
\includegraphics[width=8.5cm]{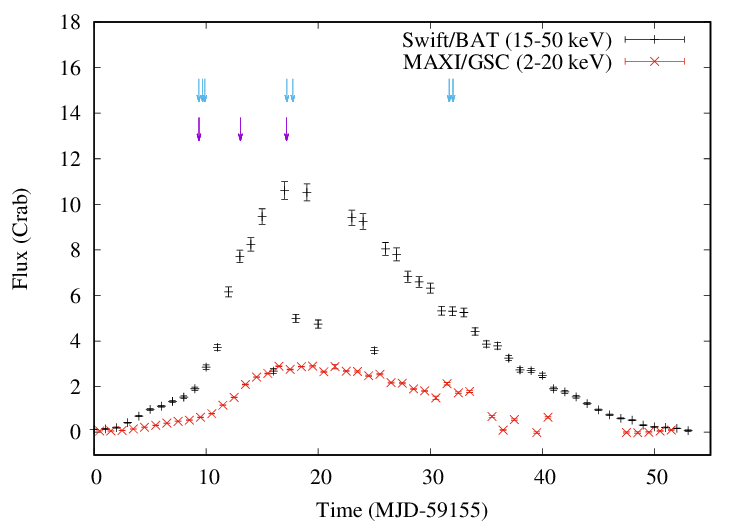}
\caption{A giant outburst detected from 1A 0535+262 using {\it Swift}/BAT and {\it MAXI}/GSC during 2020 November--December. Blue arrows show the time of {\it NuSTAR} observations, and the purple arrows show the time of {\it NICER} observations, respectively.}
	 \label{fig:BAT}}
\end{figure}

\begin{table*}
\centering
\caption{Log of {\it NuSTAR} and {\it NICER} observations.}
\label{tab:log_table}
\begin{tabular}{ccccc} 
	\hline
	Instrument & Start time       & Date & Exposure  & ObsID \\
		   &   (MJD)          & (yyyy-mm-dd)& (ks) &                    \\
\hline

       &59164.34  & 2020-11-11 & 4.94 & 90601334002 (Obs 1) \\
       & 59164.68 & 2020-11-11 & 2.61 & 90601334003 (Obs 2) \\
       & 59164.88 & 2020-11-11 & 5.16 & 90601334004 (Obs 3) \\
{\it NuSTAR} & 59172.19 & 2020-11-19 & 1.66 & 90601335002 (Obs 4) \\
       & 59172.73 & 2020-11-19 & 0.87 & 90601335003 (Obs 5) \\
       & 59186.69 & 2020-12-03 & 2.76 & 90601336002 (Obs 6) \\
       & 59187.02 & 2020-12-04 & 1.49 & 90601336003 (Obs 7) \\
\hline
        &59164.35  & 2020-11-11 & 12.07 & 3200360127 (Obs 1) \\
{\it NICER}   & 59168.05 & 2020-11-15 & ~3.24 & 3200360131 (Obs 2) \\
	& 59172.17 & 2020-11-19 & ~1.78 & 3200360135 (Obs 3) \\
\hline
	\end{tabular}
\end{table*}

\section{OBSERVATION AND DATA ANALYSIS}
\label{obs}
We analysed data taken by {\it NuSTAR} and {\it NICER} covering the entire phase of the outburst. The source went through an outburst from the first week of 2020 November and continued for nearly two months \citep{Ma20, Pa20}. We reduced data from different satellites using {\tt HEASOFT} version 6.27.2. We have used final data products (light curves) provided by {\it MAXI} \citep{Ma09} and the pulse frequency and pulsed flux evolution provided by {\it Fermi} \citep{Fi09, Me09} missions.

\subsection{{\it NuSTAR} observation}
On 2012 June 13, the Nuclear Spectroscopic Telescope Array ({\it NuSTAR}) was launched. This is the first X-ray orbital aiming telescope at energies beyond 10 keV. The observatory is made of two identical X-ray telescope modules that are co-aligned and running in a broad energy range 3--79 keV with an angular resolution of 18$^{\prime\prime}$ (full width at half maximum or FWHM) and half-power diameter (HPD) of 58$^{\prime\prime}$. Each solid-state CdZnTe pixel detector of each telescope (typically known as focal plane modules A (FPMA) and B (FPMB))  provides a spectral energy resolution of 400 eV (FWHM) at 10 keV \citep{Ha13}.  
{\it NuSTAR} performed several observations of 1A 0535+262 throughout the outburst and an observation log was given in Table \ref{tab:log_table}. The data were reduced using the {\tt NuSTARDAS pipeline} version 0.4.7 (2019-11-14) provided under {\tt HEASOFT} with latest {\tt CALDB} version. We extracted light curves and spectra of the source and background using {\tt NUPRODUCTS} scripts provided by the {\tt NuSTARDAS pipeline} from circular regions with radii of 50 and 90 arcsec,  respectively. 

\subsection{{\it NICER} observation}
The Neutron Star Interior Composition Explorer ({\it NICER}) onboard the International Space Station is a non-imaging, soft X-ray telescope. The X-ray Timing Instrument (XTI) is the main part of {\it NICER} and operates in a soft X-ray region (0.2--12 keV) \citep{Ge16}. A series of follow-up observations continued with {\it NICER} from 2020 November 11, after the {\it Swift}/BAT detection of the giant outburst. Table \ref{tab:log_table} summarized the {\it NICER} observations log used for the current study. The processing of basic data was done with {\tt NICERDAS} in {\tt HEASOFT}. We created clean event files by applying the standard calibration and filtering tool {\tt nicerl2} to the unfiltered data. We extracted light curves and spectra using {\tt XSELECT}. 
  For the timing analysis, we selected good time intervals according to the following conditions: ISS not in the South Atlantic Anomaly (SAA) region, source elevation $>$ 20$^{\circ}$ above the Earth limb, and source direction at least 30$^{\circ}$ from the bright Earth.
 For the timing analysis, we applied barycentric corrections using the task {\tt barycorr}. The ancillary response file and response matrix file of version 20200722 were used in our spectral analysis. The background corresponding to each epoch of the observation was simulated by using the {\tt nibackgen3C50} tool\footnote{\url{https://heasarc.gsfc.nasa.gov/docs/nicer/tools/nicer_bkg_est_tools.html}}\citep{Re21}. 

\subsection{Timing analysis}
We extracted the light curves from science event data in different energy ranges from the {\it NuSTAR} and {\it NICER} during the outburst. To generate the light curves in different energies, we used 10 s time binning. We selected barycentre-corrected, clean, and consistent event data to look for periodicity and used the {\tt efsearch} task in {\tt FTOOLS}. We folded the light curve over a trial period range and determined the best period by maximizing $\chi{^2}$ \citep{Le87} as a function of the period over 32 phase bins per period. Uncertainty in the estimated spin period was computed using the task {\tt efsearch} in {\tt FTOOLS} from the chi-square versus spin period plot \citep{Ra10}. After determining the best spin period from observations, pulse profiles were produced using the {\tt efold} task in {\tt FTOOLS} by folding light curves. To investigate the spin evolution of the X-ray pulsar, we used the frequency history for the source provided by the GBM pulsar archive\footnote{\url{https://gammaray.nsstc.nasa.gov/gbm/science/pulsars}}. We also compared the values of pulse frequencies and their higher derivatives during the outburst with {\it Fermi}/GBM. 

\subsection{Spectral analysis}   
 X-ray spectra were extracted using the different {\it NuSTAR} (3--79 keV) observations during the outburst. In the case of {\it NuSTAR} data, we first extracted an image of the source and then choose a suitable source region and the background region. We extracted the source and background spectra using these regions. We used {\tt XSPEC} \citep{Ar96} version 12.11.0 for spectral fitting of {\it NuSTAR} data. The spectra were rebinned to have at least 25 counts per energy bin using the task {\tt grppha}. The spectra were well described with absorbed cut-off power law and a blackbody emission component. We added a Gaussian component to model the emission line of the iron K$_{\alpha}$ fluorescence line around 6.4 keV. The cyclotron absorption feature was modelled with a multiplicative Gaussian absorption (GABS), which improved the fitting statistics of the {\it NuSTAR} spectra significantly. We kept the hydrogen column density ($N_{H}$) as a free parameter during the spectral fitting to investigate if any local soft component absorption plays an important role in specifying spectral parameter values.  

\section{RESULT}
\label{res}
In this section, we summarize the results of our timing and spectral analysis of the X-ray pulsar 1A 0535+262 during the 2020 giant outburst. The outburst from 1A 0535+262 was detected by {\it Swift}/BAT and {\it MAXI}/GSC as shown in Fig. \ref{fig:BAT}.

\subsection{Evolution of pulse period and pulse profile}
 The nature of the pulse profile bears crucial information concerning the geometric properties of the neutron star emitting region. From the timing analysis using {\it NuSTAR} observation, 1A 0535+262 showed a regular pulsation of $103.58\pm0.01$ s during the rising phase of the outburst, which gradually decreased at a rate of $\dot P= -1.75\pm 0.05\times10^{-7}$ s s$^{-1}$. The variation of the pulse period and its derivative was also consistent with the pulse period evolution as recorded with {\it Fermi}/GBM during the outburst. The evolution of pulse profiles was shown in Fig. \ref{fig:energy_dependent_pulse_profile} using different {\it NuSTAR} observations at different flux levels. The pulse profiles (3--79 keV) at higher luminosity (supercritical regime) showed multiple peaks. The pulse profiles showed high variability and strong energy dependency during the outburst. 

\begin{figure*}
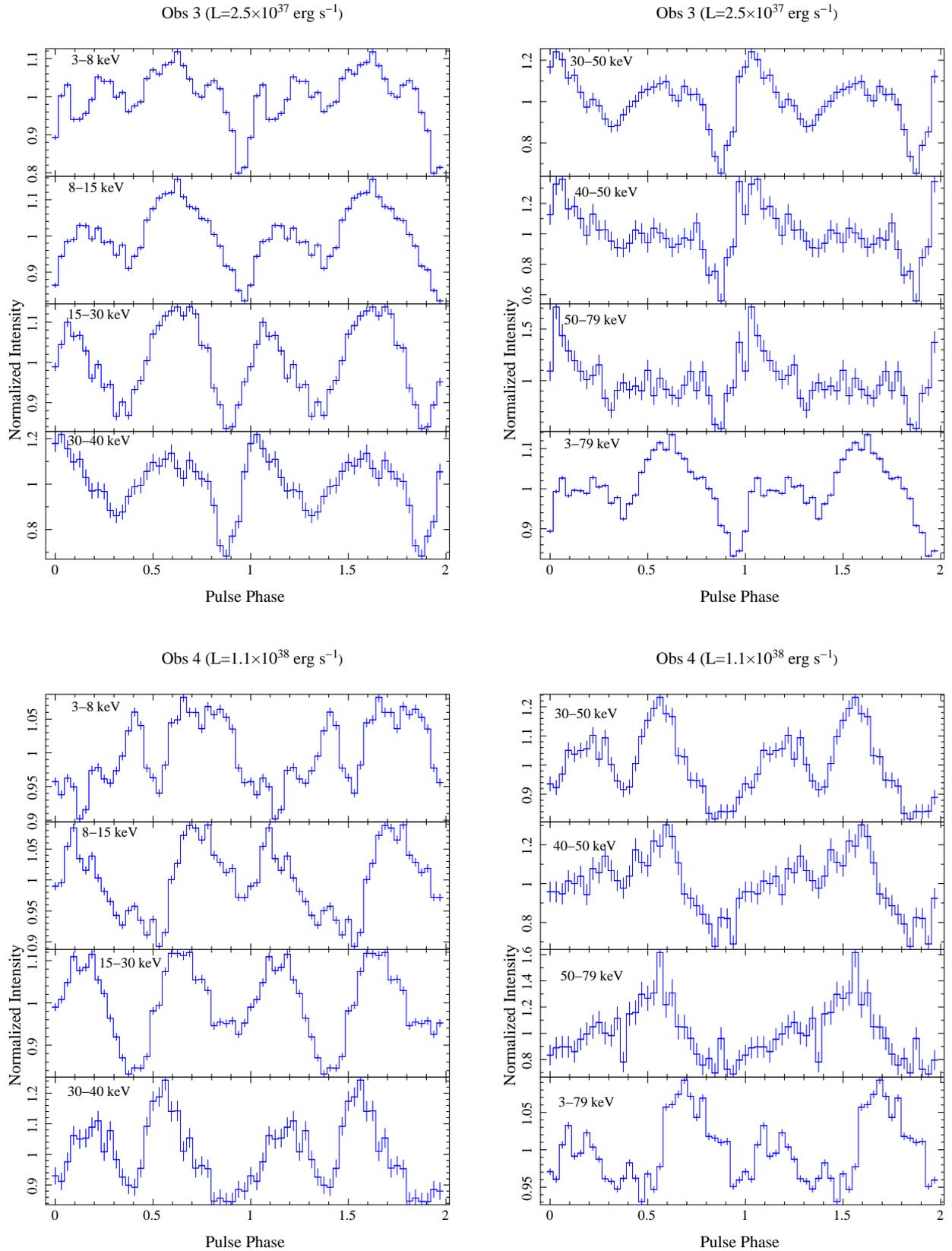

	\centering{
\includegraphics[width=8.5cm]{erp1_nu4004V2.eps}
\includegraphics[width=8.5cm]{erp2_nu4004V2.eps}
\includegraphics[width=8.5cm]{erp1_nu5002V2.eps}
\includegraphics[width=8.5cm]{erp2_nu5002V2.eps}
\caption{Energy dependent pulse profiles for different {\it NuSTAR} observations. The source showed highly variable pulse profiles with energy and luminosity. Pulse profiles evolved significantly near the energy close to CRSF.}
	\label{fig:energy_dependent_pulse_profile}}
\end{figure*}

\begin{figure*}
	\centering{
\includegraphics[width=8.5cm]{erp1_nu6002V2.eps}
\includegraphics[width=8.5cm]{erp2_nu6002V2.eps}
\includegraphics[width=8.5cm]{erp1_nu6003V2.eps}
\includegraphics[width=8.5cm]{erp2_nu6003V2.eps}\\
\contcaption{}
}
\end{figure*}

\begin{figure}
\centering{
\includegraphics[width=8.5cm]{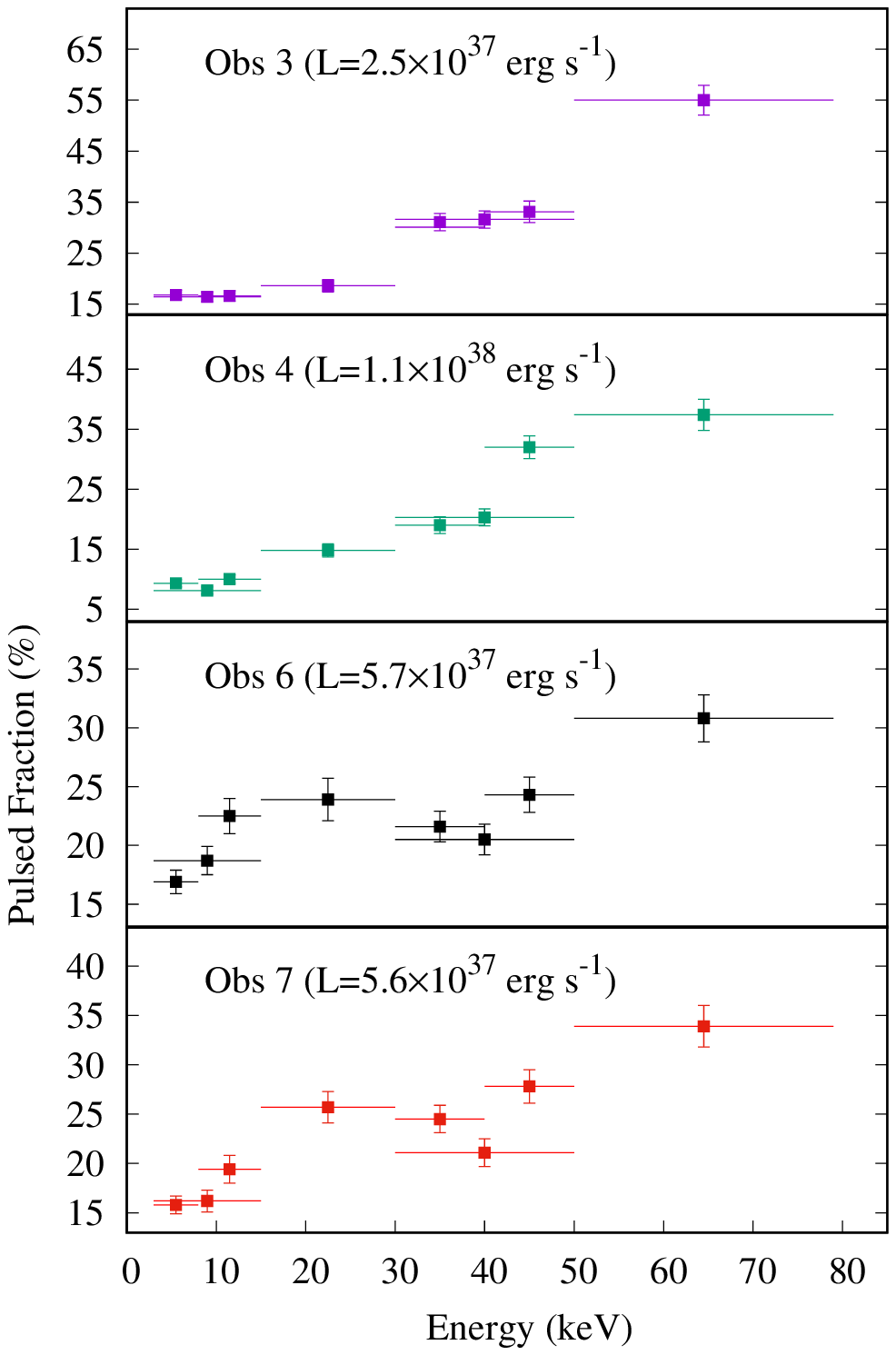}
	\caption{Variation of pulsed fraction with energy for different {\it NuSTAR} observations. The horizontal error bars indicate the energy range for which the pulsed fraction was calculated and the vertical error bars show errors of corresponding measurements. Luminosity ($L$) of each observations was estimated in the energy range 3--79 keV for a source distance of 2 kpc.} 
\label{fig:pulse_fraction}
}
\end{figure}

\begin{figure}
\centering{
\includegraphics[width=8.8cm]{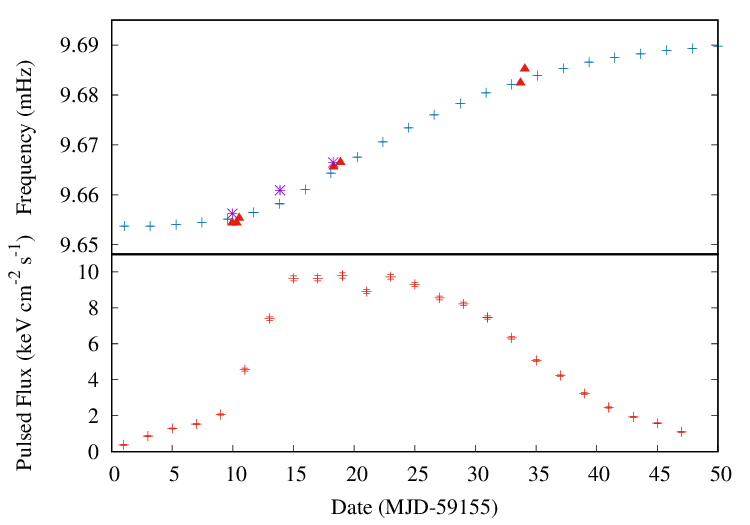}
\caption{The top panel shows the evolution of pulse frequency over the outburst. The blue points show the variation of pulse frequency using {\it Fermi}/GBM data and the purple asterisks show the pulse frequencies calculated using {\it NICER} data and the red triangles represent corresponding frequencies as estimated from {\it NuSTAR} data. The bottom panel shows the variation of pulsed flux in the energy range 12--50 keV using {\it Fermi}/GBM data.} 
\label{fig:frequency variation}}
\end{figure}

\begin{figure}
\centering{
\includegraphics[width=8.8cm]{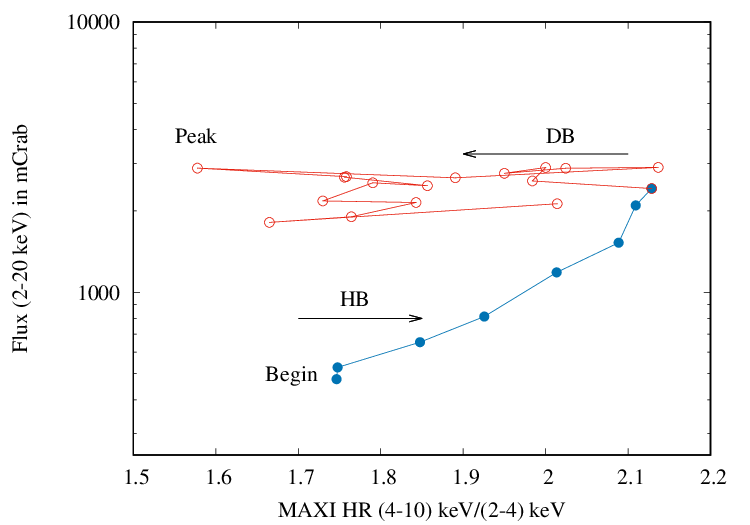}
\caption{ Hardness intensity diagram (HID) of 1A 0535+262 during MJD 59162--59186 using {\it MAXI} flux. Open red circles corresponding points in the diagonal branch (DB) and filled blue circles representing points in the horizontal branch (HB). A transition from HB to DB is visible above the critical luminosity which indicates a transition from subcritical to supercritical regime.
}
\label{fig:Hardness}}
\end{figure}

\subsection{Energy-resolved pulse profile}
The energy dependence of the pulse profiles was studied throughout the outburst using different {\it NuSTAR} observations. 
 The energy-dependent pulse profiles were shown in Fig. \ref{fig:energy_dependent_pulse_profile} for different {\it NuSTAR} observations at different luminosities. During the rising phase (obs 3), the pulse profile clearly showed the double peak feature in the lower energy, and the pulse profile evolved to a single peak in the higher energy range ($>$40 keV). A significant evolution of profile was observed near the cyclotron line energy, probably due to the change in the beaming pattern near CRSF.
Near the peak of the outburst (obs 4), the pulse profile also showed a dual-peak feature in low energy ($<$40 keV), and the profile evolved to a single peak feature in the high energy range. During the declining phase of the outburst (obs 6), a broad pulse profile was observed in low energy, and in the high energy range ($>$30 keV), the pulse profile evolved to a double peak feature close to the cyclotron line energy. For {\it NuSTAR} observation 7, in the soft X-ray region, a broad feature with dips in the pulse profile was observed. The pulse profile evolved to a double peak feature for the energy range $>$30 keV. The beaming pattern and emission geometry clearly showed significant variation near the CRSF for different observations and the beaming pattern also showed significant variation in subcritical and supercritical regimes.

We studied the variation of pulsed fraction (PF) with energy for different days of the outburst. PF is defined as the ratio between the difference of maximum intensity (${I_{\textrm max}}$) and minimum intensity (${I_{ \textrm min}}$) to their sum: [$({I_{\textrm max}}$ -- ${I_{\textrm min}})$ $/$ (${I_{\textrm max}}$ + ${ I_{\textrm min}}$)]. 
The variation of PF with energy was shown in Fig. \ref{fig:pulse_fraction}. The horizontal error bars on data points in Fig. \ref{fig:pulse_fraction} represented the energy ranges for which PF was estimated.
The pulse fraction showed a trend to increase with energy except the energy close to CRSF. An interesting feature was observed that the pulsed fraction showed a local maxima near the cyclotron line energy for multiple observations. 
  
The top panel of Fig. \ref{fig:frequency variation} showed the variation of the pulse frequency during the outburst using {\it Fermi}/GBM, which was consistent with our result. The bottom panel of Fig. \ref{fig:frequency variation} showed the evolution of pulsed flux during the outburst between 12--50 keV, which reached a record high of $\sim$10 keV cm$^{-2}$ s$^{-1}$. We also looked for the higher-order derivative of the pulse frequency ($\dot{f}$) of the source during the outburst and the estimated value of $\dot{f}$ was $\sim1.4\times10^{-11}$ Hz s$^{-1}$ using different {\it NuSTAR} and {\it NICER} observations.  

In Fig. \ref{fig:Hardness}, the hardness ratio (4--10 keV/2--4 keV) during the outburst was shown using {\it MAXI}. The hardness ratio varied from 1.6 to 2.2. 
 Fig. \ref{fig:Hardness} showed a clear transition from HB to DB during the supercritical regime which was not observed earlier for this source. Above the critical luminosity, the HID showed a sudden turn and entered to the DB, which was shown by red open circles. The blue solid points corresponded to the HB.  The HID showed a q-like feature during the outburst.
 
\subsection{Energy spectrum}
\begin{figure*}
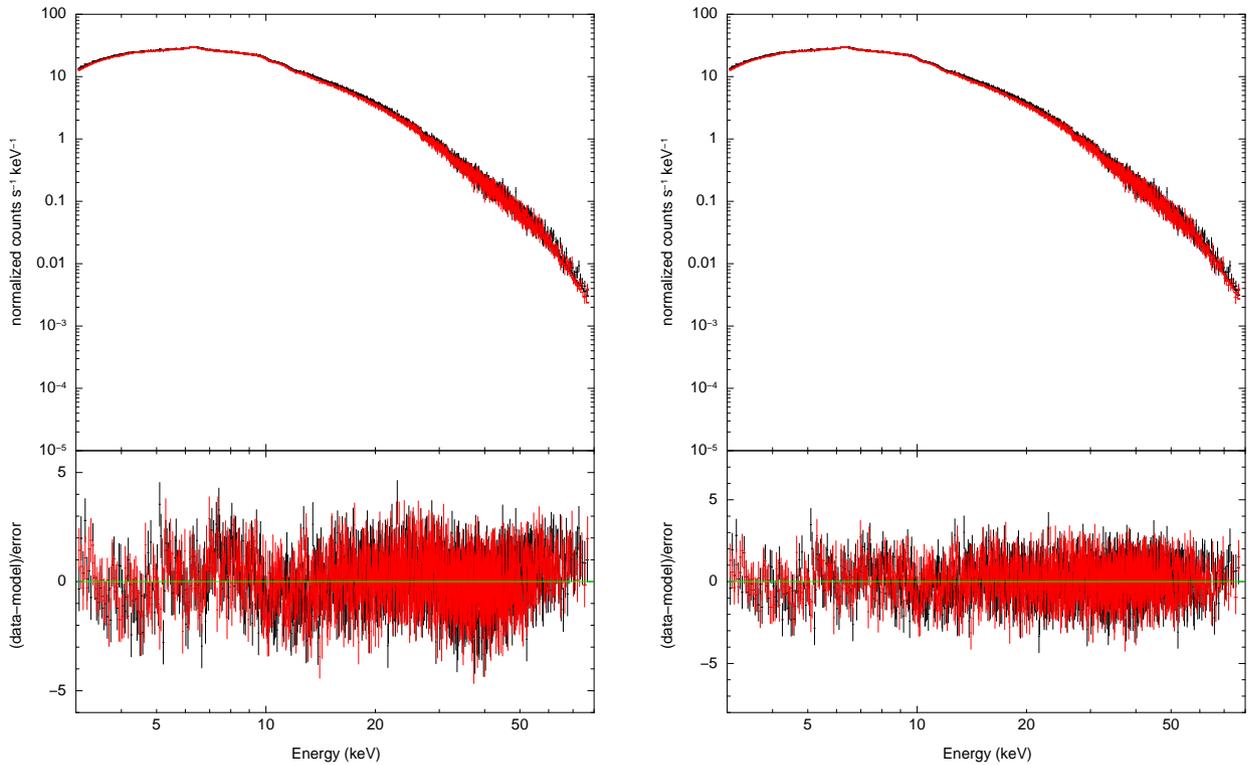

\centering{
 \includegraphics[width=8.5cm]{nu4002_grp25p_without_gabs.eps}
 \includegraphics[width=8.5cm]{nu4002_grp25p.eps}
	\caption{The left-hand side image shows the spectrum with a best-fitted model without cyclotron line component and the right-hand side image shows the spectrum with a best-fitted model with cyclotron line component (at 44 keV) along with other model components (absorbed cut-off powerlaw, a blackbody component and an iron emission line at 6.4 keV) for {\it NuSTAR} observation 1. The {\it NuSTAR}/FPMA spectrum is shown in black and {\it NuSTAR}/FPMB spectrum is shown in red. Residuals are shown in the bottom panels.}
\label{fig:spectrum}
	}
\end{figure*}

 We used {\it NuSTAR} observations for our spectral analysis since it has a wide spectral coverage and good sensitivity in the region where most cyclotron lines are found (3--79 keV).
The continuum spectra of most of the accreting X-ray pulsars are well explained by high energy cut-off power law (usually cut-off energy $E_c>10$ keV). The energy spectrum with the best-fitted model was shown in Fig. \ref{fig:spectrum}. Table \ref{tab:Spec} showed the result of different spectral fitting parameters for the best-fitted model. The energy spectrum was well fitted with an absorbed cut-off power-law, blackbody emission with a Gaussian component at $\sim$6.4 keV for the emission line of iron. The iron line was detected from all observations and the equivalent width showed a higher value during the declining phase. The CRSFs were detected in the energy range 38--46 keV with width $\sim$10 keV from {\it NuSTAR} observations, which were fitted with a Gaussian absorption model ({\tt GABS} in {\tt XSPEC}). The intraday and interday variation of spectral parameters were also studied to look for the evolution of these parameters. 

\begin{figure}
\centering{
  \includegraphics[width=8.8cm]{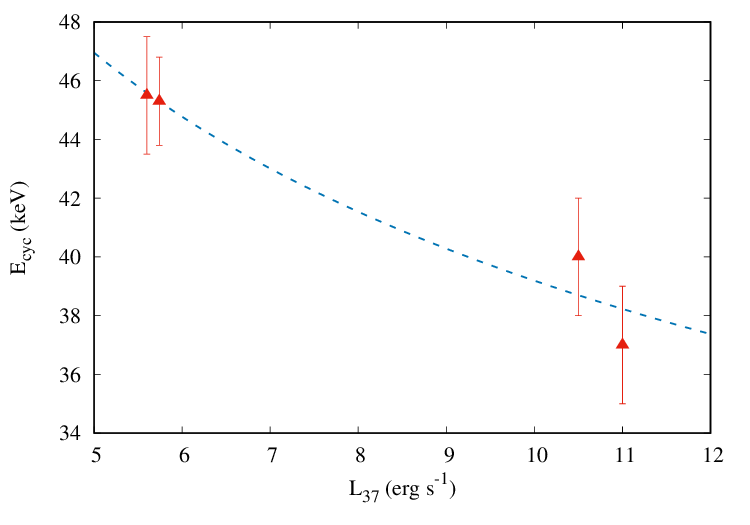}
	 \caption{A negative correlation between the cyclotron line energy and luminosity is visible in supercritical regime of the pulsar 1A 0535+262 during the giant outburst of 2020. We have plotted the points for which $L$ $\gtrsim$ $L_\textrm{crit}$.}
	 \label{fig:Eclum}}
\end{figure}

During the giant outburst of 2020, {\it NuSTAR} observed the X-ray pulsar 1A 0535+262 multiple times. The luminosity dependence of CRSFs at a high luminosity of the source was investigated.
The majority of the {\it NuSTAR} observations for 1A 0535+262 took place at a high mass accretion rate of the source.
 A clear negative correlation was found between the cyclotron line energy and luminosity as visible in Fig. \ref{fig:Eclum} during the supercritical regime.
 
\begin{figure}
\centering{
\includegraphics[width=8.8cm]{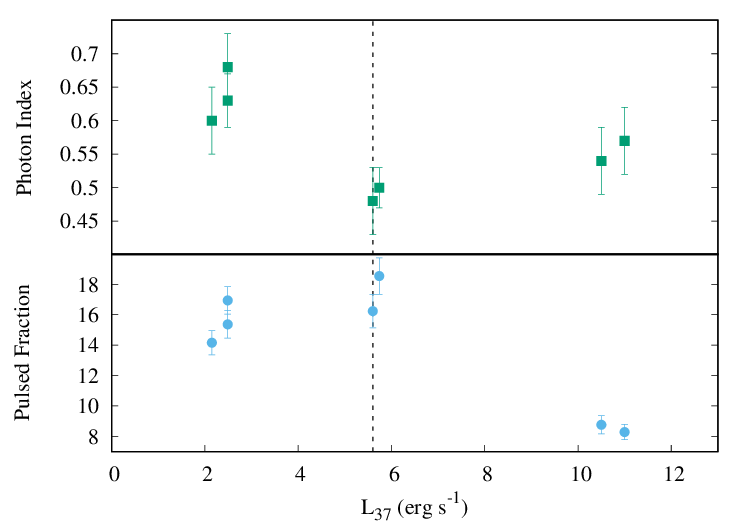}
\caption{Photon index and pulse fraction as a function of the luminosity. An abrupt change in the correlations is seen at the estimated value of the critical luminosity (shown as a vertical dotted line) at $\sim 6\times10^{37}$ erg s$^{-1}$.}
\label{fig:evo_spec}}
\end{figure}

We studied the evolution of different spectral parameters during the giant outburst. As the peak luminosity was above the critical luminosity during the outburst, a state transition from subcritical to supercritical accretion regime occurred. During this transition, different spectral parameters showed a significant variation with luminosity as the mass accretion rate also changed. We observed that the photon index showed a significant evolution in correlation with luminosity. The top panel of Fig. \ref{fig:evo_spec} showed the variation of photon index with luminosity. The source showed a positive correlation between the photon index and luminosity at a high luminosity level ($L>L_\textrm{crit}$) and a negative correlation was observed between the photon index and luminosity below the critical luminosity ($L<L_\textrm{crit}$). We estimated the turning of the correlation at the luminosity $\sim6\times 10^{37}$ erg s$^{-1}$ in the energy range 3--79 keV, assuming a distance to the source of 2 kpc \citep{St98, Ba18}.

\begin{table*}
\caption{Spectral fitting parameters for absorbed cut-off power law, blackbody emission, Gaussian component and gabs: {\tt phabs$\ast$(cutoffpl+bb+ga)$\ast$gabs} from different {\it NuSTAR} observations.}
\begin{threeparttable}
\centering
\label{tab:Spec}
\begin{tabular}{lccccccc}
\hline
	Model Parameters                &  Obs 1  &   Obs 2    & Obs 3               &  Obs 4  &  Obs 5  & Obs 6 &  Obs 7                      \\
		 &  MJD 59164.34        & MJD 59164.68 & MJD 59164.88   &        MJD  59172.19       &     MJD 59172.73         & MJD 59186.69 & MJD 59187.02  \\
\hline
	$N_{H}$ (10$^{22}$ cm$^{-2}$)  & $0.44\pm0.17$ & $0.68\pm0.2$ & ~$0.5\pm0.15$   & $0.34\pm0.15$  & ~$0.3\pm0.16$ & ~$0.3\pm0.15$  &  $0.29\pm0.13$                  \\
	$kT_{bb}$ (keV)        & $~6.3\pm0.8$ & $~6.6\pm0.8$ & $6.2\pm0.5$ & $6.3\pm0.5$ & $6.1\pm0.5$ & $6.5\pm0.9$ & $6.3\pm0.8$                \\
	norm$_{bb}$ (10$^{-1}$)                  & $~~0.9\pm0.08$ & $1.7\pm1.5$ & ~~~~~$1\pm0.09$ & ~$12\pm0.8$ & $12\pm1.1$    & $3.5\pm0.6$ & $3.7\pm0.3$                     \\  
	Photon index ($\Gamma$) & $~~0.6\pm0.05$ & $0.68\pm0.05$ & $0.63\pm0.04$  & $0.57\pm0.05$  & $0.54\pm0.05$   & ~~$0.5\pm0.05$  &  $0.48\pm0.05$                       \\
	HighECut (keV) & $17.6\pm0.8$ & $18\pm2$ & $17.7\pm0.65$ & $16.5\pm1.2$ & $14\pm2$ & $14.8\pm0.73$ & $16.3\pm0.8$                       \\
norm$_\Gamma$           & $~~0.6\pm0.04$ & $0.69\pm0.05$ & $0.68\pm0.03$ & ~~$1.7\pm0.08$ & ~$1.8\pm0.08$ & ~~~~~$1\pm0.02$  & ~~$1.1\pm0.03$ \\
	$E_{K_\alpha}$ (keV) & $6.38\pm0.02$ & $6.41\pm0.02$ & $6.33\pm0.02$ & $6.38\pm0.16$ & $6.48\pm0.02$ & $6.44\pm0.02$ & $6.38\pm0.02$        \\
	$\sigma_{K_\alpha}$ (keV) & $0.17\pm0.02$ & $0.23\pm0.03$ & $0.24\pm0.02$ & $0.36\pm0.02$  & $0.35\pm0.02$ & $0.35\pm0.02$ & $0.34\pm0.03$   \\
	  norm$_{K_\alpha}$ (10$^{-2}$)     & $~0.8\pm0.06$ & ~~~~~$1\pm0.09$ & $~~~~~1\pm0.07$ & ~~~$7\pm0.3$   & $7.1\pm0.8$  & ~~~~$3\pm0.2$ & ~~~~$3\pm0.2$ \\
	$E_{CRSF}$ (keV) & $43\pm1.5$ & $43\pm2$ & $45.5\pm1$ & $37\pm2$ & $40\pm2$ & $45.3\pm1.5$ & $45.5\pm2.5$          \\
	$\sigma_{CRSF}$ (keV) & ~$10\pm1.76$ & $11\pm3$ & $10.14\pm1.63$ & $9.2\pm2.4$ & $10.2\pm3.3$ & $11.2\pm2.4$ & ~$9.9\pm2.2$                       \\
	Flux (3--79 keV) & $4.54\times10^{-8}$ & ~~$5.2\times10^{-8}$ & ~$5.13\times10^{-8}$ & $2.26\times10^{-7}$ & $2.23\times10^{-7}$  & ~$1.27\times10^{-7}$  &   ~$1.17\times10^{-7}$                    \\
	 (erg cm$^{-2}$ s$^{-1}$) &            &           &              &    &     &                &                                 \\
	$\chi^{2}$  (d.o.f) & 1.03 (2390)  & 1.05 (2182) & 1.00 (2515)   & 1.03 (2607) & 0.99 (2347)  & 1.06 (2551)    &  1.07 (2307)                   \\
\hline
\end{tabular}
\begin{tablenotes}
\small
\item  $N_{H}$: hydrogen column density, $\Gamma$: power law photon index, $kT_{bb}$: blackbody temperature
\end{tablenotes}
\end{threeparttable}
\end{table*}

\section{DISCUSSION}
\label{dis}
We presented the timing and spectral analysis using data from different satellites during the recent giant outburst of the X-ray pulsar 1A 0535+262 in 2020. A state transition from subcritical to supercritical accretion regime was seen from the HID. We observed a transition from the HB to DB in the HID for the pulsar 1A 0535+262 first time. A similar type of transition was observed for different pulsars like 4U 0115+63, EXO 2030+375, V 0332+53, and KS 1947+300 \citep{Re13}. 1A 0535+262 entered in the DB by making a sudden turn (towards left) in the HID when the luminosity reached a critical value. Above the critical luminosity, hardness ratio (HR) started to decrease and the peak of the giant outburst corresponded to the softest state of the DB. 

From the timing analysis, the spin period of the X-ray pulsar was found $103.58\pm 0.01$ s during the rising phase of the outburst, which was consistent with the pulse period recorded with {\it Fermi}/GBM during the outburst. The pulse profiles showed time variability and strong energy dependency and the morphology of the pulse profiles evolved significantly with energy and luminosity. 
During the giant outburst of 2005, a double peak pattern was observed up to 60 keV and at higher energy one of the peaks appeared to be strongly reduced \citep{Ca07}. We also observed that the double peak feature was evolved to a single peak feature in the higher energy range, which closely matches the previous result \citep{Ca07}. 
The pulse profile evolved significantly during different phases of the outburst which implied that the emission mechanism and the beaming pattern changed significantly during the outburst.

The pulse profile showed a single peak feature in higher energy and a double peak feature in lower energy, which may suggest that the intrinsic change in the beaming pattern from a fan beam to pencil beam feature with an increase in energy. The luminosity of a pulsar influences the beaming pattern. If the luminosity of a pulsar is below the critical luminosity (subcritical regime), accreting material directly falls on the neutron star surface resulting in an X-ray emission with a `pencil beam' form.  For the `pencil beam' pattern, the accreted material reaches the neutron star surface through nuclear collisions with atmospheric protons or through coulomb collisions with thermal electrons \citep{Ha94}. For the pencil beam pattern, emission escapes from the top of the column \citep{Bu91}. When the luminosity of the source is more than the critical luminosity (supercritical regime), a radiation-dominated shock is created in the accretion column where the material slows down towards the neutron star surface and radiation pressure is high enough to stop the accreting matter at a distance above the neutron star. X-rays are released in the shape of a `fan-beam' from the sides of the accretion column in this scenario \citep{Da73}. The shock region moves higher in the accretion column, where the magnetic field is weaker, as the accretion rate increases. However, the beaming patterns can be more complex than a simple pencil or fan beam \citep{Kr95, Be12, Bl00}.

The evolution of pulsed fraction with luminosity was studied for different days of the outburst for different mass accretion rates.  The bottom panel of Fig. \ref{fig:evo_spec} showed the variation of pulsed fraction with luminosity where the dotted line indicated the critical luminosity. The figure showed an abrupt change in correlation between the pulsed fraction and luminosity near the critical luminosity. In the subcritical accretion regime ($L<L_\textrm{crit}$), the PF showed a positive correlation between the pulsed fraction and luminosity and above the critical luminosity ($L>L_\textrm{crit}$), a negative correlation was observed. This reversal in correlation implied a transition from subcritical to supercritical accretion regimes during the giant outburst. 
The observed negative correlation can be explained due to the increase in unpulsed components from the source in the pattern of fan-beam emission in the supercritical regime. When the source reaches the critical luminosity, the impact from the unpulsed photons leaving through the side walls of the column rises, which affects beam geometry and pulse morphology. 

During the 2020 giant outburst, 1A 0535+262 showed a significant change in pulse profile and pulsed fraction near the cyclotron line energy. The beaming pattern and emission geometry were affected near the cyclotron line energy due to the change in scattering cross-section near the CRSF \citep{Ar99, Sc07}. Earlier, several studies found that significant variation in pulse profile and pulsed fraction was observed near the cyclotron line energy for a few sources like 4U 0115+63 \citep{Fe11}, V 0332+53 \citep{Ts06}, 4U 1901+03 \citep{Be21}, 1A 1118--61 \citep{Lu09}, Her X--1 \citep{Lu09}, and GX 301--2 \citep{Lu09}. However, sources like 4U 1907+09, Cen X-3, Vela X-1, and 4U 1626--67 were featureless near their CRSF fundamental energies.
The increase in pulsed fraction with energy was non-uniform and has local maxima close to the fundamental cyclotron line and harmonics for the sources like V 0332+53, Vela X-1, Her X-1, GX 1+4, and 4U 0115+63 \citep{Lu09}. For Cen X--3 and 1A 0535+262, no distinct change in pulse profile was found near CRSF \citep{Lu09} which is probably due to insufficient line depth and the wide line range. 

We detected the presence of cyclotron absorption lines between 42--46 keV in the subcritical regime, which were consistent with the previous results \citep{Ke94, Mu13, Ba17, Ts19}. In the supercritical regime, the energy of the cyclotron line was found to be lower than the subcritical regime. Typically, the width of the CRSF is nearly 10 keV \citep{Sa15, Ca09} which is in good agreement with our results. Based on our results, we can calculate the value of the magnetic field of the source corresponding to the fundamental cyclotron line at $\sim$44 keV. The magnetic field of the neutron star is related to the energy of the cyclotron line as 
\begin{equation}
 E_\textrm{CRSF} = 11.57\times B_{12}(1+z)^{-1}
\end{equation}
where, $E_{\textrm CRSF}$ is the cyclotron line energy in keV, $B_{12}$ is the magnetic field in $10^{12}$ G and $z$ is the gravitational red-shift ($z\approx0.3$ for typical neutron star) \citep{Sc07, Do14, Fu14a}. We estimated the value of the magnetic field $\simeq4.9\times10^{12}$ G corresponding to the detected cyclotron line at $\sim44$ keV. A prominent emission feature was detected at $\sim6.4$ keV from the spectral analysis for different {\it NuSTAR} observations during the outburst which was associated with the fluorescent iron emission line and is often observed in the spectra of BeXRPs and consistent with previous studies \citep{Ba17, Ts19}.
 
For the sources V 0332+53 \citep{Ts06, Do17}, 4U 0115+6415 \citep{Na06}, and SMC X--2 \citep{Ja16}, a negative correlation between the $E_{\textrm{cyc}}$ and the X-ray luminosity was reported. A negative correlation is thought to be related to the change in height of the accretion column, which is supported by radiative pressure and arises only above a specific critical luminosity \citep{Ba76}.
Whether the emitting plasma's radiation pressure is capable of decelerating the accretion flow depends on the critical luminosity ($L_{\textrm{crit}}$) of a pulsar. Near the shock region, the CRSF centroid energy decreases as luminosity increases. For some sources, a positive correlation or negative correlation between luminosity and line energy was previously reported. For sources like Her X--1 \citep{St07}, GX 304--1 \citep{Kl12}, Swift J1626.6--5156 \citep{De13} and Vela X--1 \citep{Fu14b}, Cep X--4 \citep{Fu15}, a positive correlation between luminosity and line energy was reported. Critical luminosity also helps to define two accretion regimes. Luminosity is known to be correlated with the line energy below this critical luminosity, either due to a change in the atmospheric column height above the neutron star surface driven by ram pressure of the in-falling material \citep{St07} or due to the Doppler effect \citep{Mu15}.

We estimated the mass accretion rate during the outburst. The peak luminosity of the source was $\sim$1.2 $\times$ 10$^{38}$ erg s$^{-1}$, which was estimated for the energy range of 3--79 keV and distance to the source of 2 kpc. The luminosity and the mass accretion rate related as $L$ = $\eta$ $\dot{M}$ c$^{2}$. Taking $\eta$ = 0.2 \citep{Si00}, the mass accretion rate ($\dot{M}$) was estimated to be $\sim$ 6.67 $\times$ 10$^{17}$ g s$^{-1}$. 
The critical luminosity for 1A 0535+262 was estimated to be $\sim$6.78$ \times$ 10$^{37}$ erg s$^{-1}$ using theoretical calculation \citep{Be12}: 
\begin{equation}
 L_\textrm{crit} = 1.5\times 10^{37} B_{12}^{16/15} \textrm{erg s$^{-1}$}
\end{equation}
where, $B_{12}$ is the surface magnetic field in units of $10^{12}$ G.
For the pulsar 1A 0535+262, we estimated the critical luminosity to be $\sim$6 $\times$ 10$^{37}$ erg s$^{-1}$ from {\it NuSTAR} observations, which was close to the earlier estimated theoretical value \citep{Be12}. 

The photon index decreased with increasing luminosity in the subcritical accretion regime and above the critical luminosity, the photon index showed a trend to increase with luminosity. Earlier, a negative correlation was observed between the photon index and luminosity in subcritical accretion regime \citep{Mu13}. We observed a significant change in the correlation between the luminosity and photon index, suggesting that the source went through a transition between subcritical to supercritical regimes. The luminosity at which the correlation changes was identified with the critical luminosity by \citet{Re13}. Earlier, several sources showed an abrupt change in the $L$~--$~\Gamma$ diagram close to critical luminosity. The transition from a negative to positive correlation was seen in the $L$~--$~\Gamma$ diagram as luminosity increases \citep{Re13}. A negative correlation was reported for the sources like 1A 0535+262, 1A 1118--612, GRO J1008--57, XTE J0658--073 in the subcritical regime, and a transition was observed for the sources 4U 0115+63, EXO 2030+375, and KS 1947+300 \citep{Re13}. In the subcritical accretion regime, the negative correlation implied the hardening of the power law continuum with flux and in the supercritical accretion regime, the positive correlation implied the softening of the power law continuum with flux. 

\section{CONCLUSION}
\label{con}
We reported the recent giant outburst of the X-ray pulsar 1A 0535+262 which reached a record value of nearly 12.5 Crab. We summarized the result of the evolution of the timing and spectral properties of the source during the outburst. A state transition from subcritical to supercritical regime occurred during the giant outburst. The state transition was also supported by the HID. The transition from horizontal to the diagonal branch was observed from the HID. We found a q-like feature in the HID during the outburst. The pulse profile showed a highly variable nature and a strong energy dependence was established. A significant variation of pulse profile and pulse fraction was observed near cyclotron line energy. 
 The pulse frequency also changed during the outburst with a rate of $1.4\times10^{-11}$ Hz s$^{-1}$, which was comparable to the {\it Fermi}/GBM. 
The energy spectrum was described with an absorbed cut-off power law, a blackbody emission component, a Gaussian component of the iron emission line, and an additional cyclotron absorption feature was also included in the model. One of the important outcomes of this study was the detection of variable cyclotron lines with luminosity. The magnetic field of the pulsar was estimated to be $\simeq4.9\times10^{12}$ G corresponding to the cyclotron line energy at $\sim$ 44 keV. The relationship between cyclotron line parameters and luminosity is used to investigate how accretion geometry changes with the change of the accretion rate. We found that 1A 0535+262 showed a negative correlation between the cyclotron line energy and luminosity during the supercritical regime of the 2020 giant outburst. The critical luminosity for the pulsar 1A 0535+262 was estimated to be $\sim 6\times 10^{37}$ erg s$^{-1}$. The photon index and pulsed fraction also showed a significant variation during the state transition. The reversal in the correlation between the photon index with luminosity and pulsed fraction with luminosity was observed near this critical luminosity. The photon index was negatively correlated with X-ray flux in the horizontal branch but in the diagonal branch, the photon index was positively correlated with flux, which implied that as the flux increases, the spectrum becomes harder in the horizontal branch and softer in the diagonal branch.

\section*{Acknowledgements}
We thank the anonymous reviewer for his suggestions, which helps to improve the manuscript significantly. This research has made use of data obtained with {\it NuSTAR}, a project led by Caltech, funded by NASA and managed by NASA/JPL, and has utilized the {\tt NUSTARDAS} software package, jointly developed by the ASDC (Italy) and Caltech (USA). 
This research has made use of the {\it MAXI} data provided by RIKEN, JAXA, and the {\it MAXI} team. We acknowledge the use of public data from the {\it NuSTAR}, {\it NICER}, and {\it Fermi} data archives. 
\section*{Data Availability}
The data underlying this article are publicly available in the High Energy Astrophysics Science Archive Research Center (HEASARC) at \\
\url{https://heasarc.gsfc.nasa.gov/db-perl/W3Browse/w3browse.pl}.



\bibliographystyle{mnras}








\bsp	
\label{lastpage}
\end{document}